\documentclass[]{elsart}
\usepackage{graphicx,epsfig,amsmath,amsfonts,ulem}
\def\ssigma{\boldsymbol{\sigma}}

\def\vec#1{{\bf #1}}
\def\text#1{{\mathrm #1}}
\begin{document}

\begin{frontmatter}
\title{Equivalence of kinetic-theory and random-matrix approaches to Lyapunov spectra of hard-sphere systems}
\author{A. S. de Wijn$^{12}$}
\ead{A.S.deWijn@science.ru.nl}
\thanks[nwo]{The author's work is currently supported by a Veni grant of the Netherlands Organisation for Scientific Research (NWO).}
\thanks[henk]{The author would like to thank Henk van Beijeren for very stimulating discussions and his continued interest.}
\address{Institute for Theoretical Physics, Utrecht University, Leuvenlaan 4, 3584 CE, Utrecht, The Netherlands\\Max-Planck-Institut f\"ur Physik Komplexer Systeme, N\"othnitzer Stra{\ss}e 38, 01187 Dresden, Germany\\Radboud University Nijmegen, Institute for Molecules and Materials, Heyendaalseweg 135, 6525AJ Nijmegen, the Netherlands, Tel: +31(0)243652850, Fax: +31(0)243652120}
\begin{keyword}
Lyapunov exponents\sep
Lyapunov modes\sep
Goldstone modes\sep
hard spheres\sep
high-dimensional chaos\sep
random matrices\sep
\PACS {05.45.Jn} 
\end{keyword}

\begin{abstract}
\noindent
In the study of chaotic behaviour of systems of many hard spheres, Lyapunov exponents of small absolute value exhibit interesting characteristics leading to speculations about connections to non-equilibrium statistical mechanics.
Analytical approaches to these exponents so far can be divided into two groups, macroscopically oriented approaches, using kinetic theory or hydrodynamics, and more microscopically oriented random-matrix approaches in quasi-one-dimensional systems.
In this paper, I present an approach using random matrices and weak disorder expansion in an arbitrary number of dimensions.
Correlations between subsequent collisions of a particle are taken into account.
It is shown that the results are identical to those of a previous approach based on an extended Enskog-equation.
I conclude that each approach has its merits, and provides different insights into the approximations made, which include the Sto{\ss}zahlansatz, the continuum limit, and the long-wavelength approximation.
The comparison also gives insight into possible connections between Lyapunov exponents and fluctuations.
\end{abstract}

\end{frontmatter}

\section{Introduction}

In recent years, investigations into the connections between the theory of dynamical systems and non-equilibrium statistical mechanics have yielded many interesting and important results.
Gallavotti and Cohen~\cite{gc1,gc2}, for instance, conjectured that many-particle systems as studied by statistical mechanics will generally be strongly chaotic.
This has prompted a great deal of interest in the connections between chaos on the one hand and the decay to equilibrium and transport coefficients on the other (see for instance Ref.~\cite{gaspardhausdorff}).
A central role in the study of chaos and related properties is played by the Lyapunov exponents, which describe the exponential divergence or convergence of nearby trajectories in phase space.

Some of this interest has been directed towards the Lyapunov exponents of the propotype system of many hard spheres.
Several analytical calculations of, among other things, the largest Lyapunov exponent and the sum of all positive Lyapunov exponents have been performed~\cite{ramses,prlramses,cylinders,onszelf,lagedichtheid,ksentropie}.
Lyapunov exponents of many-particle systems have also been evaluated numerically in molecular-dynamics simulations (see, for instance Ref.~\cite{posch2,radons1}).
Because of their unexpected behaviour, in particular the Lyapunov exponents of small but nonzero absolute value have received attention.
A step structure occurs in the Lyapunov spectrum near zero whenever the system is large enough compared to the mean free path, as was first noted by Posch and Hirschl~\cite{posch1} and later also found in other systems (see, for example, Refs.~\cite{radons1,tnmboundary}).
These Lyapunov exponents differ from the exponents of larger absolute value, in the sense that all particles contribute to them, much like in the case of the zero Lyapunov exponents, and the corresponding modes appear to be, on average and to first approximation, linear combinations of these zero modes with a sinusoidal modulation in the position.
Initially, it was hoped that describing the Lyapunov modes through a macroscopically oriented approach such as hydrodynamics or an Enskog equation might provide insight into possible connections between chaos and transport.
In Ref.~\cite{onszelf}, it has been shown that the small exponents can in fact be viewed to belong to Goldstone modes and that the behaviour found in simulations~\cite{posch1} can be understood from this.
A set of equations was derived for these exponents by the use of an extended Enskog equation and values for the exponents were obtained.
Other attempts to understand these exponents have been based on hydrodynamic equations~\cite{mareschal}, and, although limited to quasi-one-dimensional systems, random matrices along with approximations of weak disorder~\hbox{\cite{eckmann,taniguchi1,taniguchi2}}.

In view of the two distinct approaches to the Goldstone modes, through random matrices on the one hand and through the Enskog equation on the other, it is of interest to investigate whether the results of Ref.~\cite{onszelf} can also be derived using techniques from random matrix theory.
In this paper, instead of starting from the Enskog equation, I make use of random matrices and the weak-disorder expansion.
Unlike the previous random-matrix approaches mentioned above, the present derivation is not limited to quasi-one-dimensional systems.
The approximations needed to arrive at quantitative results can be studied more carefully in some cases, and are similar to those used in the derivation of Ref.~\cite{onszelf}.
By comparing the Enskog and random-matrix approaches, one gains insight into the approximations made in both approaches and the associated inaccuracies.
Of special interest are the consequences of the thermodynamic limit, since finite-size effects in the Lyapunov exponents may be related to fluctuations and decay of correlations.

This paper is organised as follows.  In Sec.~\ref{sec:lyap}, Lyapunov exponents are briefly introduced as well as the dynamics in tangent space of freely moving hard spheres in tangent space.
Next, in Sec.~\ref{sec:goldstoneandenskog}, a summary is given of the Goldstone modes and the calculation of Ref.~\cite{onszelf} by the use of an extended Enskog equation.
In Secs.~\ref{sec:rmt}, \ref{sec:equations}, and~\ref{sec:perturbationparameter}, it is explained how the results found from the extended Enskog equation can also be derived through the use of random matrices.
The approaches are compared in Sec.~\ref{sec:comparison}, and the approximations needed are discussed.
Possible corrections are considered and it is pointed out how these may lead to insight in the connections between non-equilibrium behaviour and chaotic properties.

\section{\label{sec:lyap}\label{sec:spheresdyn}Lyapunov exponents and the dynamics in tangent space}
Consider a $d$-dimensional system of $N$ particles moving in a $2dN$-dimensional phase space $\Gamma$.
At time $t=0$, the system is assumed to be in an initial point ${\vec\gamma}_0$ in this phase space, from which it evolves with time according to ${\vec\gamma}({\vec\gamma}_0,t)$.
If the initial conditions are perturbed infinitesimally by $\delta{\vec\gamma}_0$, the system evolves along  an infinitesimally different path $\gamma(\gamma_0,t) + \delta \gamma(\gamma_0,t)$, where $\delta\gamma$ denotes a coordinate in the tangent space $\delta\Gamma$, and $\delta \gamma(\gamma_0,0)=\delta{\vec\gamma}_0$.
The evolution of a vector in the tangent space is described by
\begin{eqnarray}
{\delta{\vec\gamma}({\vec\gamma}_0,t)}
\label{eq:M}&=& {{\sf M}_{{\vec\gamma}_0}(t)\cdot \delta{\vec\gamma}_0~,}
\label{eq:matrix}
\end{eqnarray}
where ${\sf M}_{{\vec\gamma}_0}(t)$ is a ${2dN}$-dimensional matrix defined by
\begin{eqnarray}
\label{eq:tang}
{\sf M}_{{\vec\gamma}_0}(t)=\frac{d {\vec\gamma}({\vec\gamma}_0,t)}{d {\vec\gamma}_0}~.
\end{eqnarray}
The Lyapunov exponents are the possible average asymptotic growth rates of infinitesimal perturbations $\delta \gamma_i(\gamma,t)$ associated with the eigenvalues $\mu_i(t)$ of ${\sf M}_{{\vec\gamma}_0}(t)$, i.e.,
\begin{eqnarray}
\lambda_i &=&\lim_{t\rightarrow\infty} \frac{1}{t}\left(\ln|\mu_i(t)| +  i \arg \mu_i(t)\right)~.
\end{eqnarray}
If the system is ergodic, it will eventually come arbitrarily close to any point in phase space for all initial conditions except for a set of measure zero.
The Lyapunov exponents are thus the same for almost all initial conditions.
In the literature, one also finds the Lyapunov exponents defined with reference to the eigenvalues of $[{\sf M}_{{\vec\gamma}_0}(t)^\dagger\cdot{\sf M}_{{\vec\gamma}_0}(t)]^\frac{1}{2}$, in which case they are real.

The symmetries of the dynamics of the system generate vectors in tangent space which do not grow or shrink exponentially and therefore have Lyapunov exponents equal to zero.
For a system of hard spheres under periodic boundary conditions, these symmetries and their corresponding zero modes are uniform translations,
Galilei transformations, time translations, and velocity scaling.

We now consider a gas of identical hard spheres of diameter $a$ and mass $m$ in $d$ dimensions in the absense of external fields.
As there are no internal degrees of freedom, the phase space may be represented by the positions $\vec{r}_i$ and velocities $\vec{v}_i$ of all particles, enumerated by $i$, and similarly the tangent space by infinitesimal deviations $\delta\vec{r}_i$ and $\delta\vec{v}_i$.
The evolution of the system in phase space consists of a sequence of free flights interrupted by collisions.
During the free flights, the particles do not interact and their positions change linearly with the velocities; similarly, $\delta\vec{r}$ changes linearly with $\delta\vec{v}$.
For rigid spheres the collisions are instantaneous.
At the moment of the collision, momentum is exchanged between the two particles involved along the collision normal $\hat{\vec{\ssigma}} = (\vec{r}_i-\vec{r}_j)/{a}$ at impact, as shown in Fig.~{\ref{fig:bolletje}.
At the instant of the collision, none of the other particles are assumed to interact.

From Eq.~(\ref{eq:tang}) and the phase-space dynamics, the dynamics in tangent space can be derived~\cite{ramses,soft2}.
During the free flight between the instant of a collision $t_z$ ($z$ being the number of the collision in the sequence) and $t$, there is no interaction between the particles and the components of the tangent-space vector transform according to
 \begin{eqnarray}
 \left(
 \begin{array}{c}
 \delta\vec{r}_i\\
 \delta\vec{v}_i
 \end{array}
 \right)_t
 &=& {\mathcal Z}(t-t_z)\cdot \left(
 \begin{array}{c}
 \delta\vec{r}_{i}\\
 \delta\vec{v}_{i}
 \end{array}
 \right)_{t_z}~,\\
 {\mathcal Z}(t-t_z)
 &=&
 \left(\begin{array}{cc}
 \extracolsep{1mm}
 \rule{0mm}{0mm}{\sf I}&(t-t_z){\sf I}\\
 \rule{0mm}{5mm}0&{\sf I}
 \end{array}\right)
 ~,
 \label{eq:flight}
 \end{eqnarray}
 in which ${\sf I}$ is the $d\times d$ identity matrix.

\begin{figure}
\includegraphics[width=7cm]{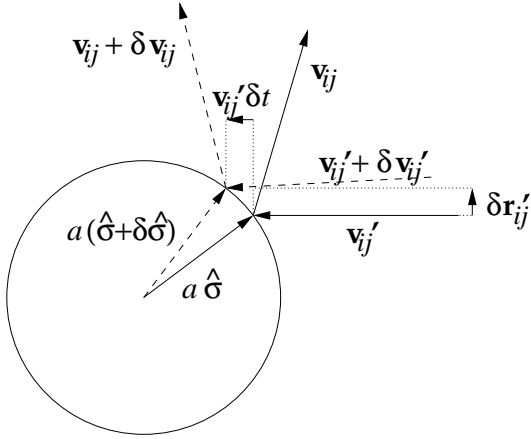}
\caption{\label{fig:bolletje} Geometry of a collision of two particles $i$ and $j$ of diameter $a$, in relative position $\vec{r}_{ij} = \vec{r}_i - \vec{r}_j$ and with relative velocity $\vec{v}_{ij} = \vec{v}_i - \vec{v}_j$.
The collision normal $\hat{\vec{\ssigma}}$ is the unit vector pointing from the centre of particle $j$ to the centre of particle $i$.
The circle drawn represents the locus of closest approach. 
Coordinates before the collision are marked with a prime.
The dashed lines indicate an infinitesimally displaced path.
}
\end{figure}

As shown in Fig.~\ref{fig:bolletje}, infinitesimal differences in the positions and velocities of the particles lead to infinitesimal changes in the collision normal and collision time.
This, in turn, leads to infinitesimal changes in both positions and velocities right after the collision.
Throughout the paper, primes denote coordinates in phase space and tangent phase space just before a collision while non-primed quantities refer to coordinates just after the collision.
For colliding particles $i$ and $j$, one finds
\begin{eqnarray}
\label{eq:mathcalL}
\left(\begin{array}{c}
\delta\vec{r}_i\\
\delta\vec{r}_j\\
\delta\vec{v}_i\\
\delta\vec{v}_j
\end{array}\right)
= ({\mathcal L}+{\mathcal I}) 
\cdot\left(\begin{array}{c}
\delta\vec{r}_i'\\
\delta\vec{r}_j'\\
\delta\vec{v}_i'\\
\delta\vec{v}_j'
\end{array}\right)  
\label{eq:collision}
&=&
\left(
\begin{array}{cccc}
\extracolsep{2mm}
\rule[-3mm]{0mm}{7mm}{\sf I-S}&{\sf S}&\rule[-5mm]{0mm}{0mm}0&0\\
\rule[-3mm]{0mm}{5mm}{\sf S}&{\sf I-S}&0&0\\
\rule[-3mm]{0mm}{7mm}{\sf -Q}&{\sf Q}&{\sf I-S}&{\sf S}\\
\rule[-3mm]{0mm}{5mm}{\sf Q}&{\sf -Q}&{\sf S}&{\sf I-S}
\end{array}\right)
\cdot\left(\begin{array}{c}
\delta\vec{r}_i'\\
\delta\vec{r}_j'\\
\delta\vec{v}_i'\\
\delta\vec{v}_j'
\end{array}\right)  
~,
\end{eqnarray}
where ${\mathcal I}$ and ${\sf I}$ are the $4d\times 4d$ and $d\times d$ identity matrices, respectively, and ${\mathcal L}$ is the $4d\times4d$ collision matrix, which can be written in terms of $d \times d$ matrices ${\sf S}$ and ${\sf Q}$ specifying the collision dynamics in tangent space (see, for instance, Refs.~\cite{ramses,onszelf}).

Let ${\sf Z}(t)$ be the $2 d N \times 2 d N$ matrix which performs the single-particle transformations ${\mathcal Z}(t)$ for all particles during free flight according to Eq.~(\ref{eq:flight}).
Let ${\sf L}_{z}$ be the $2 d N \times 2 d N $ matrix performing the transformations associated with collision number $z$ of two particles at time $t_z$, as written in Eq.~(\ref{eq:collision}), while leaving the other particles unaffected.
${\sf M}_{\gamma_0}(t)$ in Eq.~(\ref{eq:M}) then is a product of these matrices for the sequence of collisions $1,2,\dots,z$ occurring between time $t_1=0$ and $t$ and the free flights in between the collisions.
Its specific form reads
 \begin{eqnarray}
 {\sf M}_{\gamma_0}(t) = {\sf Z}(t-t_{z})\cdot{\sf L}_{z}\cdot{\sf Z}(t_{z}-t_{z-1})\cdot{\sf L}_{z-1}
 \cdot\ldots\cdot{\sf Z}(t_{2}-t_{1})\cdot{\sf L}_{1}~.
 \label{eq:product}
 \end{eqnarray}
In principle, ${\sf M}_{{\vec\gamma}_0}(t)$ in Eq.~(\ref{eq:M}) can now be derived for finite times for any finite system with given initial conditions, after which the eigenvalues of ${\sf M}_{\gamma_0}(t)$ can be determined.

\section{Goldstone modes and the extended Enskog equation\label{sec:goldstoneandenskog}\label{sec:boltzmann}}

At this point,
in order to be able to compare the random-matrix approach described in this paper to our previous analytical results obtained by using an extended Enskog equation~\cite{onszelf}, it is useful to review the derivation of the extended Enskog equation and the approximations made therein.

In Ref.~\cite{onszelf}, it was shown that the sinusoidal modes found in the simulations may be interpreted as being Goldstone modes.
These occur in systems with a continuous symmetry, such as the symmetries associated with the zero modes.
Translation invariance, for instance, causes the evolution operator to commute with the translation operator.
As a result, the two operators have a set of common eigenfunctions, which are sinusoidal perturbations of the continuous symmetry that grow or shrink slowly with time.
For these modes to stand out among the continuum of other Lyapunov exponents, their wavelength must be large compared to the typical length scale of the mean free path.  That is,
\begin{eqnarray}
\label{eq:kklein} \bar{v} k \ll \bar{\nu}~,
\end{eqnarray}
where $\bar{v}$ is the average absolute value of the velocity of the particles, and $\bar{\nu}$ is the average single-particle collision frequency.
The modes found in simulations are linear combinations of the Goldstone modes for wave vectors of equal size belonging to left- and right-moving waves.
The qualitative behaviour seen in the simulations~\cite{posch1} can be understood entirely from this.
Also, the inconsistencies in the propagation velocity and amplitude noted in Ref.~\cite{christina} can be completely understood from the behaviour of linear combinations of left- and right-moving Goldstone modes.

The mode corresponding to a particular Lyapunov exponent depends, in general, on the initial conditions of all particles in a way that is far too complicated to allow exact specification.
To find the exponents from analytical calculations, therefore, one has to resort to a statistical approximation.
In our earlier paper~\cite{onszelf}, this was accomplished by making assumptions similar to the Sto{\ss}zahlansatz in the Boltzmann equation.
The system is described not as a large number of separate particles, but rather by a distribution function of position, velocity, etc, which can be used to obtain the average behaviour.
The pre-collision pair distribution functions are approximated by a product of independent one-particle distributions, which is valid at low densities.
In the Boltzmann equation, in addition to the Sto{\ss}zahlansatz, the two one-particle distributions are evaluated at the same position $\vec{r}$.
The Enskog equation is a heuristic generalisation of this, in which the pair-distribution is approximated by the product of two one-particle distribution functions evaluated at the actual positions of the two particles, multiplied by a factor $\chi_{\text E}(n)$ equal to the
equilibrium pair correlation function at contact evaluated as a function of the density $n$ at that point.
More details on the Boltzmann equation and the Enskog theory of dense gases relevant to this study
may be found in Refs.~\cite{hcb,chapmancowling}.

To describe the dynamics in tangent space, we have previously derived~\cite{onszelf} a generalised Enskog equation for the single-particle distribution function $f(\vec{r},\vec{v},\delta\vec{r},\delta\vec{v},t)$ for the coordinates of a particle $(\vec{r}, \vec{v})$ and the tangent space vectors $\delta\vec{r}$ and $\delta\vec{v}$.
The latter are described by a single-particle distribution function which, in the thermodynamic limit, depends smoothly on position, velocity, and time, just like the velocity distribution in the ordinary Enskog equation.
If in addition the distribution functions of the tangent space 
vectors of two particles about to collide are assumed to factorise in a way similar to the distribution of their 
velocities, one ends up with a generalised Enskog equation that includes the tangent space variables.
If the tangent space variables $\delta \vec{r}$ and $\delta\vec{v}$ are integrated over, this equation reduces to the standard Enskog equation.

Because $\delta\vec{r}$ and $\delta\vec{v}$ are infinitesimal, the dynamics in tangent space are linear in these quantities.
Therefore, from the extended Enskog equation, one may obtain closed linear equations for the time evolution of the average first moments $\delta\vec{r} (\vec{r},\vec{v}, t)$ and $\delta\vec{v} (\vec{r},\vec{v}, t)$,
which are the averages of the tangent space vectors $\delta\vec{r}_i$ and $\delta\vec{v}_i$ of the particles 
in a small region around position $\vec{r}$ and velocity $\vec{v}$ at time $t$.
The result is a set of linear equations for the averages~\cite{onszelf}, reading
\begin{eqnarray}
\frac{\partial}{\partial t} \delta \vec{r}(\vec{r},\vec{v}, t)
 & = & - \vec{v}\cdot \frac\partial{\partial \vec{r}} \delta\vec{r}(\vec{r},\vec{v}, t)
+
\label{eq:boltzr}
 \delta\vec{v}(\vec{r},\vec{v}, t) + {\sf C}_{\sf S}\delta \vec{r}(\vec{r},\vec{v}, t)~,\\
\frac{ \partial}{ \partial t} \delta \vec{v} (\vec{r},\vec{v}, t)
& = & - \vec{v}\cdot \frac{\partial}{\partial \vec{r}} \delta \vec{v}(\vec{r},\vec{v}, t)
+ {\sf C}_{\sf S} \delta \vec{v}(\vec{r},\vec{v}, t) + {\sf C}_{\sf Q} \delta \vec{r}(\vec{r},\vec{v}, t)~.
\label{eq:boltzv}
\end{eqnarray}
The linear functional collision operators ${\sf C}_{\sf S}$ and ${\sf C}_{\sf Q}$ are associated with the matrices ${\sf S}$ and ${\sf Q}$, and specified by the collision integrals for a particle with outgoing velocity $\vec{v}$ over all outgoing velocities $\vec{u}$ for the other particle,
\begin{eqnarray}
{\sf C}_{\sf S} \, \delta\vec{q} (\vec{r},\vec{v}, t)  = \lefteqn{\int_{\hat{\vec{\ssigma}} \cdot (\vec{v}-\vec{u})\leq 0}\, d\vec{u}\, d\hat{\vec{\ssigma}} \, \chi_{\mathrm E}(n)n a^{d-1}\, |\,\hat{\vec{\ssigma}} \cdot (\vec{v}-\vec{u})|\phi_{\mathrm{M}}(\vec{u})}\nonumber\\
\label{eq:CS}\phantom{\int}&& \times \left\{ \,\delta\vec{q} (\vec{r},\vec{v}', t) + {\sf S} \cdot \left[ \,\delta\vec{q} (\vec{r} + a \hat{\vec{\ssigma}},\vec{u}', t)-\,\delta\vec{q} (\vec{r},\vec{v}', t)\right] 
\right.\nonumber\\ && \phantom{\int \times \{}\left.\null
  -\,\delta\vec{q} (\vec{r},\vec{v}, t) \right\}~,\\
{\sf C}_{\sf Q} \, \delta \vec{r}(\vec{r},\vec{v}, t)  = \lefteqn{ \int_{\hat{\vec{\ssigma}} \cdot (\vec{v}-\vec{u})\leq 0}\, d\vec{u}\, d\hat{\vec{\ssigma}}
\, \chi_{\mathrm E}(n)n a^{d-1}\, |\,\hat{\vec{\ssigma}} \cdot (\vec{v}-\vec{u})|\phi_{\mathrm{M}}(\vec{u})}\nonumber\\
\label{eq:CQ}\phantom{\int}&& \times {\sf Q} \cdot \left[ \delta\vec{r} (\vec{r} + a \hat{\vec{\ssigma}},\vec{u}', t) -  \,\delta\vec{r} (\vec{r},\vec{v}', t)\right] ~,
\end{eqnarray}
where $\delta\vec{q}(\vec{r},\vec{v}, t)$ denotes either $\delta\vec{r}(\vec{r},\vec{v}, t)$ or $\delta\vec{v}(\vec{r},\vec{v}, t)$.
As in Sec.~\ref{sec:spheresdyn} and Fig.~\ref{fig:bolletje}, the primes denote the variables before the collision with collision normal $\hat{\ssigma}$ and outgoing velocities $\vec{v}$ for one particle and $\vec{u}$ for the other, while $\phi_{\mathrm{M}}(\vec{u})$ is the Maxwell distribution.

The equations for the first moments can be solved by using spatial Fourier transforms.
The solutions have the form
\begin{eqnarray}
\delta\vec{q}_i = \vec{f}_\vec{k}(\vec{v}_i) \exp{({\mathrm i} \vec{k}\cdot\vec{r}_i+\lambda t)}~,\label{eq:genformsol}
\end{eqnarray}
where $\vec{k}$ is the wave vector.
For the Lyapunov exponents there is reasonable quantitative correspondence to the results of simulations~\cite{onszelf}.
Surprisingly, it turned out that the Sto{\ss}zahlansatz still affects the leading order in the density~\cite{onszelf,proefschrift}.

The extended Enskog equation, the equation for the first moments, and its solutions are discussed in more detail in Ref.~\cite{onszelf}.

\section{\label{sec:rmt}Products of random matrices and the weak disorder expansion}

Because Eq.~(\ref{eq:product}) contains a product of similar matrices, a more natural approach to the problem of Lyapunov exponents in many-particle systems may be to consider the tangent-space maps as products of random matrices. 
Some assumptions regarding the randomness of the generating dynamics must be made to replace the matrices by a random ensemble, similarly to the assumptions used in, for example, Refs.~\cite{cylinders,benettin1984,palvul1986,onslorentz}).
The matrices in the ensemble must be symplectic, because of the time-reversal symmetry of the dynamics, yet satisfy quite strong additional restrictions and correlations between their elements.
In the case of a general many-particle system, it is not directly obvious how to deal with these remaining correlations, as pointed out in Ref.~\cite{cylinders}.

Attempts to describe the Goldstone modes through random-matrix theory have been made by Eckmann and Gat~\cite{eckmann}, and by Taniguchi, Dettmann, and Morriss~\cite{taniguchi1,taniguchi2}.
These works considered quasi-one-dimensional systems, in which particles cannot move past each other, and any particle collides only with its two neighbours.
As a result, the calculations are very much simplified, and indeed fail to produce quantitatively reliable results for fully two- and three-dimensional systems.
It should be noted in this context that for the same reason quasi-one-dimensional systems do not satisfy the Sto{\ss}zahlansatz, which does not hold if pairs of particles that have already collided collide again.
Additionally, these calculations make use of approximations that are quite similar to the weak-disorder expansion, while the applicability of the weak-disorder expansion is uncertain.
In the present section, I will indicate how the requirements of the weak-disorder expansion correspond to the assumptions needed to derive the extended Enskog equation and the equations for the first moments in the tangent space of a single particle, Eqs.~(\ref{eq:boltzr}) and~(\ref{eq:boltzv}).

For the problem under consideration the matrix product ${\sf M}_{\gamma_0}(t)$ is given in Eq.~(\ref{eq:product}).
The matrices can be taken together in pairs corresponding to a collision $j$, ${\sf L}_j$, and following free flight, ${\sf Z}(t_{j+1}-t_{j})$, to produce a product of correlated sparse random matrices
\begin{eqnarray}
\label{eq:Bproduct}
{\sf M}_{\gamma_0}(t_{z+1}) &=& {\sf B}_z\cdot{\sf B}_{z-1}\cdot\ldots\cdot{\sf B}_1~,\\
{\sf B}_j&=&{\sf Z}(t_{j+1}-t_{j}) \cdot {\sf L}_j ~.
\end{eqnarray}
After a collision, following a velocity-dependent free-flight time, each particle will be involved in another collision, with the outgoing velocity of the previous collision as the new incoming velocity.
The matrices describing such collisions are correlated.

In the next two paragraphs, the weak-disorder expansion is introduced, along with the conditions that must be met.
In the remainder of the section, it is shown how these requirements can be met for the product in Eq.~(\ref{eq:Bproduct})
and how the Lyapunov exponents can be derived.

In preparation for this, let us consider a quite general product of $z$ matrices of the form
\begin{eqnarray}
\label{eq:productsimple}{\sf M} &=& {\sf A}_z\cdot{\sf A}_{z-1}\cdot\ldots\cdot{\sf A}_1~,\\
\label{eq:A}{\sf A}_j &=& \bar{{\sf A}} + \epsilon {\sf X}_j~,
\end{eqnarray}
in which $\epsilon$ is a small perturbation parameter and $\bar{{\sf A}}$ is the average matrix, assumed to be diagonalisable.
The matrix ${\sf M}$ thus expands to
\begin{eqnarray}
{\sf M}
&= & \bar{{\sf A}}^z + \epsilon \sum_j \bar{{\sf A}}^{z-j}\cdot {\sf X}_j \cdot \bar{{\sf A}}^{j-1} + \ldots~.
\end{eqnarray}
The terms of higher orders in $\epsilon$ are not necessarily smaller than the first-order terms, because the number of terms grows faster than the inverse of their size for $z\rightarrow \infty$.
However, if the ${\sf X}_i$ are independent of one another this problem disappears and the sum converges~\cite{drieitalianen,zanonderrida}.

Let ${\sf X}_j$ be independent and identically distributed, while every element has zero average.
Derrida, Mercheri, and Pichard~\cite{derrida87} have derived that the growth rates $\lambda_i$ of the eigenvalues of the product can to first order in $\epsilon$ be expressed in the eigenvalues $\kappa_i$ of the average matrix $\bar{{\sf A}}$, provided these eigenvalues exist and are non-degenerate.
More specifically,
\begin{eqnarray}
\lambda_i &=& \ln{\kappa_i}
+ \epsilon^2 \left(\frac{\overline{(\vec{a}_i\cdot {\sf X} \cdot \vec{a}_i)^2}}{2 \kappa_i^2} - \sum_{j=1}^{i} \frac{\overline{(\vec{a}_i\cdot {\sf X} \cdot \vec{a}_j)\, (\vec{a}_j\cdot {\sf X} \cdot \vec{a}_i)}}{\kappa_i\kappa_j}\right)\nonumber\\
&&\null+ \epsilon^3 \left( \sum_{j,l=1}^i \frac{\overline{(\vec{a}_i\cdot {\sf X} \cdot \vec{a}_j)\,(\vec{a}_j\cdot {\sf X} \cdot \vec{a}_l)\,(\vec{a}_l\cdot {\sf X} \cdot \vec{a}_i)}}{\kappa_i \kappa_j\kappa_l}
\right.\nonumber\\
&&\phantom {\null+\epsilon^3(j}\left.
- \sum_{j=1}^i \frac{\overline{(\vec{a}_i\cdot {\sf X} \cdot \vec{a}_j)\,(\vec{a}_j\cdot {\sf X} \cdot \vec{a}_i)\,(\vec{a}_i\cdot {\sf X} \cdot \vec{a}_i)}}{\kappa_j \kappa_i^2}+ \frac{\overline{(\vec{a}_i\cdot {\sf X} \cdot \vec{a}_i)^3}}{3\kappa_i^3}\right)\nonumber\\
&&\null+\ldots~.
\label{eq:wde}
\end{eqnarray}
Here, $\vec{a}_i$ is the eigenvector of $\bar{{\sf A}}$ belonging to the eigenvalue $\kappa_i$.
This theorem was later extended to the degenerate case by Zanon and Derrida~\cite{zanonderrida}.
In the following sections, it will be shown that the eigenvalues of ${\sf A}$ in the present problem can be chosen arbitrarily, to be distinct, but approaching unity, so as to simplify the derivation.
For more details on random matrices and the weak-disorder expansion, see Crisanti, Paladin, and Vulpiani~\cite{drieitalianen}.

In order to apply the weak-disorder expansion to the problem under consideration, the matrix product ${\sf M}_{\gamma_0}(t)$ in Eq.~(\ref{eq:Bproduct}) must first be rewritten into the form of Eq.~(\ref{eq:productsimple}), i.e., in such a way that the matrices become independent and that an expansion parameter $\epsilon$ as introduced in Eq.~(\ref{eq:A}) exists.

Instead of calculating the spectrum of Lyapunov exponents of the original product ${\sf M}_{\gamma_0}(t)$, the spectrum of a very similar product will therefore be considered, namely
\begin{eqnarray}
{\sf O}_z^{-1} \cdot{\sf M}_{\gamma_0}(t) \cdot{\sf O}_0 =
{\sf O}_z^{-1} \cdot{\sf B}_z \cdot {\sf O}_{z-1} \cdot {\sf O}_{z-1}^{-1}\cdot{\sf B}_{z-1}\cdot\ldots\cdot{\sf B}_1\cdot{\sf O}_0~,
\label{eq:Uprod}
\end{eqnarray}
where ${\sf O}_j$ may depend on the entire phase space and may be chosen in any suitable way.
Without loss of generality, ${\sf O}_0$, may be taken to be unitary,
and the other ${\sf O}_j$ are chosen in such a way that the $2dN\times2dN$ matrices ${\sf A}_j$, defined by
\begin{eqnarray}
{\sf A}_j =  {\sf O}_j^{-1} \cdot{\sf B}_j \cdot {\sf O}_{j-1}~,
\label{eq:U}
\end{eqnarray}
satisfy the requirements of the weak-disorder expansion.
The matrices ${\sf O}_j$ are transformations to another orthogonal basis just before and back after every collision and free flight, while the matrices ${\sf A}_j$ are the tangent space maps in these new bases.  The basis vectors make up the columns of the matrix $ {\sf O}_j$.

In Sec.~\ref{sec:equations}, I derive an equation for the components of these vectors, and consequently for the elements of ${\sf O}_j$, from the requirements of the weak-disorder expansion.
These basis vectors may depend on the position of the system in phase space, i.e., the positions and velocities of the particles.
The matrices ${\sf O}_j$ need not be unitary.
Any kind of invertible transformation
will do, as long as the transformation of a product of tangent space maps is equal to the product of the transformed maps.
Here, however, I restrict myself to linear transformations, because they suffice for the present purpose.

The Lyapunov spectrum of the product ${\sf O}_z^{-1} \cdot{\sf M}_{\gamma_0}(t) \cdot {\sf O}_{0}$, Eq.~(\ref{eq:Uprod}), can then be related to the Lyapunov spectrum of the original product ${\sf M}_{\gamma_0}(t)$.
As ${\sf O}_j$ is a transformation from one orthogonal basis to another, it can be written as a product of a diagonal matrix ${\sf D}_j$ and a unitary matrix ${\sf U}_j$,
\begin{eqnarray}
{\sf O}_j = {\sf U}_j \cdot {\sf D}_j~.
\label{eq:Uprod2}
\end{eqnarray}
The first diagonal matrix ${\sf D}_0$ is the $2dN\times 2dN$ identity matrix.
The Lyapunov spectrum of ${\sf D}_z^{-1} \cdot  {\sf U}_z^{-1} \cdot {\sf M}_{\gamma_0}(t) \cdot {\sf O}_{0}$ can be determined using the Osledec theorem, which states that the growth of an arbitrary $i$ dimensional volume is dominated by the $i$ largest Lyapunov exponents.
The unitary transformations ${\sf U}_z^{-1}$ and ${\sf O}_{0}$ do not affect the volume mapped by the product, and so the growth of a volume mapped by ${\sf U}_z^{-1} \cdot {\sf M}_{\gamma_0}(t) \cdot {\sf O}_{0}$ is the same as that of a volume mapped by ${\sf M}_{\gamma_0}(t)$.
Let $\omega_i$ be the $i$-th largest eigenvalue of ${\sf D}_z^{-1}$.
One can write ${\sf D}_z^{-1}$ as a product of $z$ diagonal matrices and apply the Osledec theorem to this product as well.
The Lyapunov exponents of this product are equal to $\ln(\omega_i)/z$.
By starting from the volume mapped by ${\sf U}_z^{-1} \cdot {\sf M}_{\gamma_0}(t) \cdot {\sf O}_{0}$ and applying the Osledec theorem to its map by ${\sf D}_z$, one obtains that the growth rate of an $i$ dimensional volume mapped by the entire product in Eq.~(\ref{eq:Uprod}), is equal to the sum of the largest $i$ Lyapunov exponents of each of the two products ${\sf M}_{\gamma_0}(t)$ and ${\sf D}_{z}^{-1}$.
The Lyapunov exponents $\tilde \lambda_i$ of the complete product in Eq.~(\ref{eq:Uprod}) can therefore be written as
\begin{eqnarray}
\tilde \lambda_i = \lambda_i + \frac{\ln\omega_i}{z}~.
\label{eq:tildelambda}
\end{eqnarray}

\section{Basis vectors for the transformations to independent matrices\label{sec:equations}}

In the previous section, a framework was set up which allows one to implement the weak-disorder expansion.
In order to materialise this, time-dependent matrices ${\sf O}_j$ must be constructed that make the random matrices ${\sf A}_j$, Eq.~(\ref{eq:U}), occurring in the product Eq.~(\ref{eq:Uprod}) independent.
The columns of ${\sf O}_j$ are made up of a time-dependent set of $2dN$-dimensional basis vectors $\vec{o}_x(t)$, enumerated by $x$ which runs from $1$ to $2dN$.
It should be noted that it is convenient to choose vectors $\vec{o}_x(t)$ which are orthogonal.

In the many-particle system under consideration, the correlation between the matrices derives from the fact that information is carried between collisions through the position and velocity of a particle.
The basis vectors in which to express the tangent space maps ${\sf A}_j$
must be chosen in such a way that this correlation is removed.
In this section, this is achieved by focussing on the subsequent collisions of one particle.
Most correlation with collisions of other particles can be removed by using the Sto{\ss}zahlansatz.
The remaining correlation, which involves only a small number of particles can be used to determine a set of equations for $\vec{o}_x(t)$.

\begin{figure}
\epsfig{figure=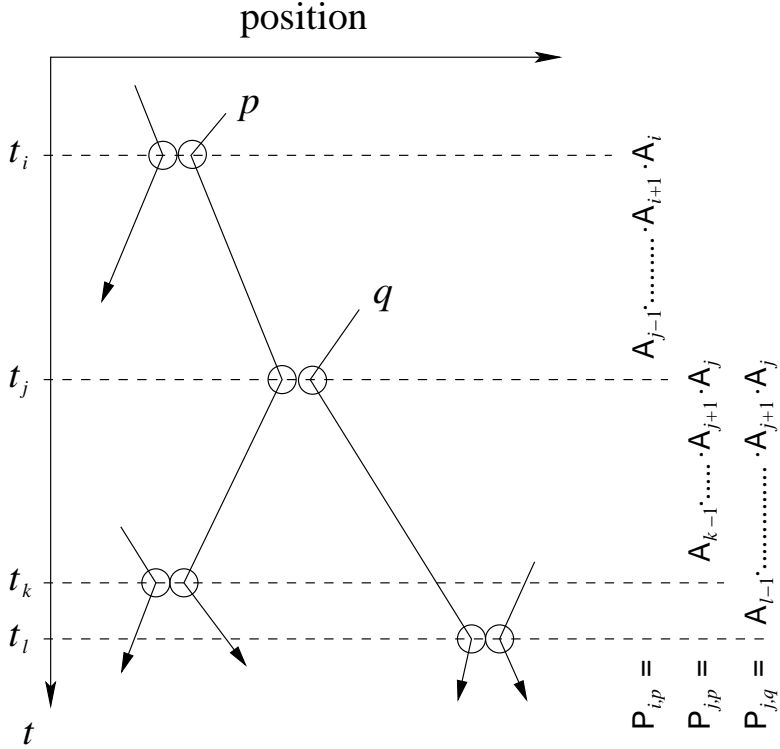,angle=90,height=10cm}
\caption{
\label{fig:diagrammetje}
Schematic representation of the collisions and free flights of particle $p$ and the particles it interacts with at collisions $i$, $j$, $k$, and, indirectly, $l$, at times $t_i$, $t_j$, $t_k$, and $t_l$, along with the corresponding tangent space maps and their products.
Because of the Sto{\ss}zahlansatz, the only remaining correlation between random matrices in the product is due to the information carried by particles between collisions.
If the matrices ${\sf A}_i,~\ldots,~{\sf A}_{l-1}$ are constructed in such a way as to make ${\sf P}_{i,p}$ independent of ${\sf P}_{j,p}$ and ${\sf P}_{j,q}$, then this correlation is eliminated.
}
\end{figure}

Let ${\sf P}_{i,p}$ be the product of matrix ${\sf A}_{j-1}\cdot\ldots\cdot{\sf A}_{i+1}\cdot{\sf A}_i$ which describes the dynamics in tangent space of all the particles at collision $i$ involving particle $p$ and during the subsequent free flight of particle $p$ until it collides with particle $q$ in collision $j$ (see Fig.~\ref{fig:diagrammetje}).
Using Eq.~(\ref{eq:U}), the $x,y$ element of ${\sf P}_{i,p}$ can be written in terms of the basis vectors $\vec{o}_x(t_j^-)$ and $\vec{o}_y(t_i^-)$ as
\begin{eqnarray}
({\sf P}_{i,p})_{xy} = \frac{\vec{o}_x(t_j^-)}{|\vec{o}_x(t_j^-)|^2} \cdot {\sf B}_{j-1} \ldots {\sf B}_{i} \cdot \vec{o}_y(t_i^-)~,
\label{eq:Pelement}
\end{eqnarray}
where the superscript in $t_j^-$ etc. denotes a time an infinitesimally small interval before the collision at that time.
Because of the Sto{\ss}zahlansatz, any collisions of particles with no common history with $p$ or $q$ are not correlated with collisions of $p$ or $q$.
As our purpose here is to remove all correlation, and these collisions are already uncorrelated, they do not need to be considered explicitely.

In order to remove the remaining correlation, and for the matrices ${\sf A}_j$ to become fully independent and to satisfy the requirements of the weak-disorder expansion, one must consider the remaining correlation, which is due to correlations between particles which share a common history.
Though there will be correlation between collisions of such particles with other particles, the corresponding tangent space maps ${\sf A}_j$ should be independent.
For this, it is sufficient that the product ${\sf P}_{i,p}$ be independent of both ${\sf P}_{j,p}$ and ${\sf P}_{j,q}$.
This ensures that the matrix products are independent of the history of a particle before its most recent collision, and consequently that collisions of any other particles which share history with $p$ and $q$ can be ignored.
The component of $\vec{o}_x(t)$ belonging to a specific particle should not depend on anything but its history since its most recent collision.
Just before collision $j$, therefore, only the recent history of particle $p$ can lead to correlation.
Consequently, only contributions to $({\sf P}_{i,p})_{xy}$ from the projections onto the tangent space of particle $p$ can lead to correlations with ${\sf P}_{j,p}$ and ${\sf P}_{j,q}$, i.~e., we may restrict ourselves to the term
\begin{eqnarray}
\left(\frac{{\mathcal P}_p\cdot \vec{o}_x(t_j^-)}{|\vec{o}_x(t_j^-)|^2}\right) \cdot {\mathcal P}_p \cdot {\sf B}_{j-1} \ldots {\sf B}_{i}
\cdot \vec{o}_y(t_i^-)
~,
\label{eq:Pelement2}
\end{eqnarray}
where ${\mathcal P}_p$ denotes the projection of the $2dN$-dimensional tangent space onto the $2d$-dimensional tangent space of particle $p$.
Furthermore, we may write
\begin{eqnarray}
\vec{o}_x(t) &=& \left(\delta\vec{r}^{(x)}_1(\vec{r}_1,\vec{v}_1,t),\delta\vec{v}^{(x)}_1(\vec{r}_1,\vec{v}_1,t)\right.,
\nonumber\\
& &
\left. \ldots, \delta\vec{r}^{(x)}_N(\vec{r}_N,\vec{v}_N,t),\delta\vec{v}^{(x)}_N(\vec{r}_N,\vec{v}_N,t)\right)~,
\label{eq:ux}
\end{eqnarray}
where
$\delta\vec{r}^{(x)}_1(\vec{r},\vec{v},t)$ $\ldots$ $\delta\vec{v}^{(x)}_N(\vec{r},\vec{v},t)$ are $d$-dimensional vector functions which are different for every~$x$.
As the particles are identical, the functions $\delta\vec{r}^{(x)}_1(\vec{r},\vec{v},t),$ $\delta\vec{r}^{(x)}_2(\vec{r},\vec{v},t)$, etc are the same for all particles, as are $\delta\vec{v}^{(x)}_1(\vec{r},\vec{v},t),$ $\delta\vec{v}^{(x)}_2(\vec{r},\vec{v},t)$, etc., so that the indices can be dropped and they can simply be
writen as $\delta\vec{r}^{(x)}(\vec{r},\vec{v},t)$ and $\delta\vec{v}^{(x)}(\vec{r},\vec{v},t)$.

As the only information carried by particle $p$ between two collisions is its position $\vec{r}_p$ and velocity $\vec{v}_p$, the product ${\sf P}_{i,p}$ is independent of  ${\sf P}_{j,p}$ and ${\sf P}_{j,q}$ if ${\sf P}_{i,p}$ is independent of the position and the velocity of the particle $p$ between the two collisions $i$ and $j$.
To this end, the functions $\delta\vec{r}^{(x)}(\vec{r},\vec{v},t)$ and $\delta\vec{v}^{(x)}(\vec{r},\vec{v},t)$ must be chosen in such a way that ${\sf P}_{i,p}$ becomes independent of $\vec{r}_p$ and $\vec{v}_p$.
For this it is necessary that at least the average $\langle{\sf P}_{i,p}\rangle$ is independent of the velocity of particle $p$ after collision $i$,
where the average $\langle . \rangle$ is calculated over all parameters, including the velocity of all particles that particle $p$ has collided with, but excluding the position and velocity of particle $p$ after collision $i$.
If this condition is met, averages of products of first moments of the matrices ${\sf P}_{i,p}$ with those of ${\sf P}_{j,p}$ and ${\sf P}_{j,q}$ behave as if the matrices are independent,
\begin{eqnarray}
\langle{\sf P}_{i,p} {\sf P}_{j,p}\rangle &=& \langle{\sf P}_{i,p}\rangle \langle {\sf P}_{j,p}\rangle ~,\\
\langle{\sf P}_{i,p} {\sf P}_{j,q}\rangle &=& \langle{\sf P}_{i,p}\rangle \langle {\sf P}_{j,q}\rangle ~.
\end{eqnarray}
The condition of independent first moments is a necessary prerequisite for the independence requirement of the weak-disorder expansion, if not necessarily sufficient.
It is assumed here that the remaining correlation can be removed by further refining the maps ${\sf O}_j$ [see Eq.~(\ref{eq:U})], if necessary through the use of nonlinear maps.
As only the higher moments of the matrices would be involved in this, it will not produce extra contributions to the leading order in the perturbation parameter in Eq.~(\ref{eq:wde}).

One must consider the relevant contribution to $\langle {\sf P}_{i,p}\rangle$ from particle $p$ in Eq.~(\ref{eq:Pelement2}), which is proportional to
\begin{eqnarray}
\left\langle\left(\delta\vec{r}^{(x)}(\vec{r}_p,\vec{v}_p,t_j^-), \delta\vec{v}^{(x)}(\vec{r}_p,\vec{v}_p,t_j^-)\right) \cdot {\mathcal P}_p \cdot {\sf B}_{j-1} \ldots {\sf B}_{i} \cdot \vec{o}_y(t_i^-)\right\rangle ~,
\end{eqnarray}
where ${\mathcal P}_p$ denotes the projection of the $2dN$-dimensional tangent space onto the $2d$-dimensional tangent space of particle $p$.
The functions $\delta\vec{r}^{(x)} (\vec{r}_p,\vec{v}_p,t)$ and $ \delta\vec{v}^{(x)} (\vec{r}_p,\vec{v}_p,t)$ do not contain any of the quantities which are averaged over.
Therefore the quantity
\begin{eqnarray}
\left(\delta\vec{r}^{(x)}(\vec{r}_p,\vec{v}_p,t_j^-), \delta\vec{v}^{(x)}(\vec{r}_p,\vec{v}_p,t_j^-)\right) \cdot \langle {\mathcal P}_p \cdot {\sf B}_{j-1} \ldots {\sf B}_{i} \cdot \vec{o}_y(t_i^-)\rangle
\label{eq:sillyproduct}
\end{eqnarray}
must be independent of $\vec{r}_p$ and $\vec{v}_p$.
If the average has the same functional dependence (up to an arbitrary prefactor) on the position and velocity of particle $p$ just before collision $i$ as $\delta\vec{r}^{(x)}(\vec{r}_p,\vec{v}_p,t_j^-)$, then the innerproduct in Eq.~(\ref{eq:sillyproduct}) is equal to an arbitrary constant
times the norm of $\mbox{\boldmath $($}\delta\vec{r}^{(x)}(\vec{r}_p,\vec{v}_p,t), \delta\vec{v}^{(x)}(\vec{r}_p,\vec{v}_p,t)\mbox{\boldmath $)$}$ squared.
In this case, ${\sf P}_{i,p}$ represented in the two bases sets just before collisions $j$ and $i$ is independent of the position and velocity of particle $p$ between the two collisions.

Consequently, one arrives at an eigenvalue-type equation for the tangent-space basis vectors with a strictly positive, but otherwise arbitrary, scalar prefactor ${\mathcal D}$, which may be different for each combination of basis vectors.
The collision parameters must be averaged out, and their distribution is the same as in Enskog theory.
Let
$\delta\vec{r}(\vec{r},\vec{v},t)$ and $\delta\vec{v}(\vec{r},\vec{v},t)$
be shorthand for
$\delta\vec{r}^{(x)}(\vec{r}_p,\vec{v}_p,t)$ and $\delta\vec{v}^{(x)}(\vec{r}_p,\vec{v}_p,t)$,
the eigenvalue equation reads
\begin{eqnarray}
\lefteqn{
\left(\begin{array}{c}
\delta\vec{r}\mbox{\boldmath $($}\vec{r} ,\vec{v},t\mbox{\boldmath $)$}\\
\delta\vec{v}\mbox{\boldmath $($}\vec{r} ,\vec{v},t\mbox{\boldmath $)$}
\end{array}\right)
=
}&
\nonumber\\
&
 {\mathcal D}
\int_{\hat{\vec{\ssigma}} \cdot (\vec{v}-\vec{u})\leq 0}
d\vec{u}\,
d\hat{\ssigma}\,
d\tau \,
\exp[{-\tau\bar{\nu}(\vec{v})}] \chi_{\mathrm {E}}(n) a^{d-1}|\,\hat{\ssigma}\cdot(\vec{v}-\vec{u})| \phi_{\mathrm{M}}(\vec{u})
\nonumber\\
&
\times
{\mathcal Z}\mbox{\boldmath $($}\tau\mbox{\boldmath $)$}
\cdot
\left(\begin{array}{cccc}
1&0&0&0\\
0&0&1&0
\end{array}\right)
\cdot
({\mathcal L}_{j}+{\mathcal I})\cdot
\left(\begin{array}{c}
 \delta\vec{r}(\vec{r}-\tau\vec{v},\vec{v}',t-\tau) \\ \delta\vec{r}(\vec{r}-\tau\vec{v}+a\hat{\ssigma},\vec{u}',t-\tau) \\ \delta\vec{v}(\vec{r}-\tau\vec{v},\vec{v}',t-\tau) \\ \delta\vec{v}(\vec{r}-\tau\vec{v}+a\hat{\ssigma},\vec{u}',t-\tau)
\end{array}\right)~,
\label{eq:groteintegraal}
\end{eqnarray}
where the collision normal is $\hat{\ssigma}$, and the incoming velocities $\vec{v}'$ and $\vec{u}'$ of two particles are mapped to the outgoing velocities $\vec{v}$ and $\vec{u}$, respectively.
The $4d\times4d$ matrix ${\mathcal L}_{j}$ represents the collision dynamics at collision $j$ as given in Eq.~(\ref{eq:mathcalL}) and the $2d\times 2d$ matrix ${\mathcal{Z}}(\tau)$ defined in Eq.~(\ref{eq:flight}) describes the free-flight dynamics.
The time between the two collisions of the particle is denoted by $\tau$.
The velocity-dependent average collision frequency is represented by $\bar{\nu}(\vec{v})=1/\bar{\tau}(\vec{v})$, where $\bar{\tau}(\vec{v})$ is the average mean free time as a function of the particle velocity.
The various solutions of Eq.~(\ref{eq:groteintegraal}) correspond to different basis vectors $\vec{o}_x(t)$, which should be orthogonal.

The strictly positive time-dependent prefactor ${\mathcal D}$ controls the eigenvalues of $\bar{\sf A}$ as well as the growth rate of the eigenvalues of ${\sf D}_j$ in Eq.~(\ref{eq:Uprod2}).
In principle, it can be chosen in any arbitrary way, as long as it is strictly positive, as any growth can be accounted for either in the basis vectors or $\bar{\sf A}$.  The choice of ${\mathcal D}$ does not affect the final result for the Lyapunov exponents, as can be seen from Eq.~(\ref{eq:tildelambda}).
If ${\mathcal D}$ is taken to be unity, $\bar{\sf A}$ is equal to the identity matrix, and the eigenvalues of $\bar{\sf A}$ are all equal to unity.
All growth will be accounted for in the time-dependence of the functions $\delta\vec{r}(\vec{r},\vec{v},t)$ and $\delta\vec{v}(\vec{r},\vec{v},t)$.
The weak-disorder expansion, however, has only been proven for non-degenerate systems and systems with doubly degenerate eigenvalues~\cite{zanonderrida}.
I therefore choose ${\mathcal D}$ differently for every basis vector, so that the eigenvalues of $\bar{\sf A}$ become distinct and the weak-disorder expansion holds.
Furthermore, ${\mathcal D}$ is chosen so close to unity that terms proportional to ${\mathcal D}-1$ become negligible compared to other terms.
This greatly simplifies the calculation.

Let ${\sf T}$ be the operator $\int_0^\infty d\tau \, \bar{\nu}(\vec{v}) \exp[{-\tau\bar{\nu}(\vec{v})}]$.
By substituting Eqs.~(\ref{eq:CS}) and (\ref{eq:CQ}) into Eq.~(\ref{eq:groteintegraal}),
one finds
\begin{eqnarray}
\frac{1}{\mathcal D}
\delta\vec{r}\mbox{\boldmath $($}\vec{r},\vec{v},t \mbox{\boldmath $)$}
&=&
{\sf T}\left\{\left[\bar{\tau}(\vec{v}){\sf C}_{\sf S}+ 1 +{\tau}{\bar{\tau}(\vec{v})} {\sf C}_{\sf Q}\right] \delta\vec{r}(\vec{v},\vec{r}-\tau \vec{v},t-\tau)
\right.\nonumber\\ &&\phantom{\sf T}\left.\null+
\tau [\bar{\tau}(\vec{v}) {\sf C}_{\sf S}+1 ] \delta\vec{v}(\vec{v},\vec{r}-\tau \vec{v},t-\tau)\right\}~,
\\
\frac{1}{\mathcal D}
\delta\vec{v}\mbox{\boldmath $($}\vec{r} ,\vec{v},t\mbox{\boldmath $)$}
&=&
{\sf T}\left\{ {\bar{\tau}(\vec{v})}  {\sf C}_{\sf Q} \delta\vec{r}(\vec{v},\vec{r}-\tau\vec{v},t-\tau)
\nonumber\right.\\ &&\phantom{\sf T}\left.\null+
[{\bar{\tau}(\vec{v})}{\sf C}_{\sf S}+1] \delta\vec{v}(\vec{v},\vec{r}-\tau \vec{v},t-\tau)\right\}~.
\end{eqnarray}
The operator ${\sf T}$ equals the unity operator when working on functions that do not depend on $\tau$, and therefore one may multiply the left side of the equations by it.
As both sides then contain the same integral over ${\tau}$, a change of variables may be performed, namely $\vec{r}' = \vec{r} - \tau \vec{v},~ t' = t-\tau$.
After renaming the variables the equations become
\begin{eqnarray}
\frac{1}{\mathcal D}
{\sf T}\delta\vec{r}\mbox{\boldmath $($}\vec{r} + \tau \vec{v},\vec{v},t + \tau\mbox{\boldmath $)$}
&=&
[\bar{\tau}(\vec{v}) {\sf C}_{\sf S}+ 1] \delta\vec{r}(\vec{v},\vec{r},t)
\nonumber\\
&& \hskip-3\bigskipamount \null
+ {\sf T} \tau \bar{\tau}(\vec{v}) [{\sf C}_{\sf Q} \delta\vec{r}(\vec{v},\vec{r},t)
+ ({\sf C}_{\sf S}+1 ) \delta\vec{v}(\vec{v},\vec{r},t)]~,\label{eq:taur}\\
\frac{1}{\mathcal D}
{\sf T}\delta\vec{v}\mbox{\boldmath $($}\vec{r}  + \tau \vec{v},\vec{v},t+ \tau\mbox{\boldmath $)$}
&=&{\tau}(\vec{v})   {\sf C}_{\sf Q}  \delta\vec{r}(\vec{v},\vec{r},t)  + [{\tau}(\vec{v}) {\sf C}_{\sf S}+1] \delta\vec{v}(\vec{v},\vec{r},t)~.\label{eq:tauv}
\end{eqnarray}

In the continuum limit the mean free time and the mean free path become small in comparison with the typical scales of the solutions.
The quantities \hbox{$\delta\vec{r}(\vec{r} - \tau \vec{v},\vec{v},t - \tau)$} and \hbox{$\delta\vec{v}(\vec{r} - \tau \vec{v},\vec{v},t - \tau)$} can thus be expanded around $\delta\vec{r}(\vec{r},
\vec{v},t)$ and \hbox{$\delta\vec{r}(\vec{r},\vec{v},t)$}, and all terms of higher order in $\tau$ can be neglected.
After substitution of Eq.~(\ref{eq:tauv}) into Eq.~(\ref{eq:taur}) and performing the integration over $\tau$, one finds
\begin{eqnarray}
\lefteqn{
\frac{1}{\mathcal D}
{\left(\bar{\tau}(\vec{v}) \vec{v}\cdot\frac{\partial}{\partial \vec{r}} + \bar{\tau}(\vec{v}) \frac{\partial}{\partial t} +1 \right)\delta\vec{r}(\vec{r},\vec{v},t)}
} \phantom{nnnn}&& \nonumber\\
&=& [\bar{\tau}(\vec{v}){\sf C}_{\sf S}+ 1] \delta\vec{r}(\vec{v},\vec{r},t)+\bar{\tau}(\vec{v}) \delta\vec{v}(\vec{v},\vec{r},t)~,\\
\lefteqn{
\frac{1}{\mathcal D}
\left(\bar{\tau}(\vec{v}) \vec{v}\cdot\frac{\partial}{\partial \vec{r}} + \bar{\tau}(\vec{v}) \frac{\partial}{\partial t} +1 \right)\delta\vec{v}(\vec{r},\vec{v},t)
}\phantom{nnnn}&&\nonumber\\
&=& {\bar{\tau}(\vec{v})}  {\sf C}_{\sf Q}  \delta\vec{r}(\vec{v},\vec{r},t)  + [{\bar{\tau}(\vec{v})}{\sf C}_{\sf S}+1] \delta\vec{v}(\vec{v},\vec{r},t)~.
\end{eqnarray}
Rearranging the terms and dividing the equations by $\bar{\tau}(\vec{v})$, followed by taking the limit of ${\mathcal D} \rightarrow 1$, leads to
two equations very similar to Eqs.~(\ref{eq:boltzr}) and (\ref{eq:boltzv}),
\begin{eqnarray}
\frac{\partial}{\partial t} \delta \vec{r}(\vec{r},\vec{v}, t)
 & = & - \vec{v}\cdot \frac\partial{\partial \vec{r}} \delta\vec{r}(\vec{r},\vec{v}, t)
+
\label{eq:boltzrD}
 \delta\vec{v}(\vec{r},\vec{v}, t) + {\sf C}_{\sf S}\delta \vec{r}(\vec{r},\vec{v}, t)\nonumber\\
&&\null+ ({\mathcal D} -1) \left[ \left({\sf C}_{\sf S} + \frac{1}{\bar\tau(v)}\right)\delta\vec{r}(\vec{r},\vec{v}, t) +\delta\vec{v}(\vec{r},\vec{v}, t)\right]
~,\\
\frac{ \partial}{ \partial t} \delta \vec{v} (\vec{r},\vec{v}, t)
& = & - \vec{v}\cdot \frac{\partial}{\partial \vec{r}} \delta \vec{v}(\vec{r},\vec{v}, t)
+ {\sf C}_{\sf S} \delta \vec{v}(\vec{r},\vec{v}, t) + {\sf C}_{\sf Q} \delta \vec{r}(\vec{r},\vec{v}, t)\nonumber\\
&&\null+
({\mathcal D} -1) \left[
{\sf C}_{\sf Q}\delta\vec{r}(\vec{r},\vec{v}, t) +
\left({\sf C}_{\sf S} + \frac{1}{\bar\tau(v)}\right)\delta\vec{v}(\vec{r},\vec{v}, t)
\right]
~.
\label{eq:boltzvD}
\end{eqnarray}
As ${\mathcal D}$ has been chosen arbitrarily close to unity, and the other terms are nonzero, the terms containing ${\mathcal D} -1$ are negligible.
We are then left with equations identical to Eqs.~(\ref{eq:boltzr}) and (\ref{eq:boltzv}).
The latter were derived from the Enskog equation by using the same approximations in a different order.
The equations can be Fourier transformed and have solutions of the form of Eq.~(\ref{eq:genformsol}).
The solutions give the components of the basis vectors that make up ${\sf O}_i$ in Eq.~(\ref{eq:U}).
As the average of the transformed matrices, $\bar{\sf A}$, is also arbitrarily close to unity, to leading order the Lyapunov exponents are simply equal to the growth rates of the basis vectors, the solutions of Eqs.~(\ref{eq:boltzr}) and~(\ref{eq:boltzv}).

\section{Perturbation parameter\label{sec:perturbationparameter}}

The pursuit for independently distributed matrices ${\sf A}_j$ has yielded the same equations as were found in Ref.~\cite{onszelf}.
However, in order for ${\sf A}_j$ to fully satisfy the requirements of the weak-disorder expansion, and thus for the Lyapunov exponents to equal the $\lambda$ found from Eq.~(\ref{eq:boltzr}) and~(\ref{eq:boltzv}), ${\sf A}_j$ must be of the form of Eq.~(\ref{eq:A}).
A small perturbation parameter $\epsilon$ must exist, i.e.,
the higher-order corrections in Eq.~(\ref{eq:wde}) must be small compared to the leading order.

The solutions to Eqs.~(\ref{eq:boltzr}) and (\ref{eq:boltzv}), which are of the form given by Eq.~(\ref{eq:genformsol}), describe the vectors that make up the columns of the matrix ${\sf O}_b$ associated with a specific collision $b$ at time $t_b$ and the subsequent free flight..
We may estimate the elements $\epsilon {\sf X}^{ij}_b$ of the corresponding disorder matrix by considering the inner product of these basis vectors given by Eq.~(\ref{eq:genformsol}), with the tangent-space collision-matrix and free-flight matrix in between.
Here $i$ and $j$ determine not only the wave vectors, but also the zero mode modulated by them.
The contribution from particle $l$ to the tangent space vector at time $t_b$ can be written as~\cite{onszelf}
\begin{eqnarray}
\label{eq:basis1}
\delta\vec{q}^{(i)}(\vec{r}_l,\vec{v}_l,t_b) &=& [\Delta\vec{q}^{(i)}(\vec{v}_l)]\exp({\mathrm{i}} \vec{k}_i \cdot \vec{r}_l + \lambda_i t_b)~,\\
\Delta\vec{q}^{(i)}(\vec{v}_l) &=& \Delta_0\vec{q}^{(i)}(\vec{v}_l) + O\left(\frac{\bar{v} k_i}{\bar{\nu} N}\right)~,
\label{eq:basis2}
\end{eqnarray}
where $\Delta_0\vec{q}^{(i)}$ denotes a linear combination of the zero modes that belongs to the mode of $\lambda_i$.
Note that the components of $\Delta_0^{(i)}\vec{q}(\vec{v}_l)$ are of order $1/N$ due to normalisation.
If $k_i$ becomes large, then, because of normalisation, the components are of order $1/N<k_i/N$.
Note that modes with different wave vectors $\vec{k}_i$ decouple on average after taking the continuum limit.

For all particles in the time interval $\Delta t_b \sim 1/(\bar{\nu} N)$ between this collision $b$, and the next one of the entire system, $b+1$, the position changes by a small amount $\vec{v}_l \Delta t_b$.
More explicitly,
the matrix ${\sf A}_b$ related to any specific collision $b$ and the directly following free flight [see Eq.~(\ref{eq:Uprod})] can be expressed as
\begin{eqnarray}
({\sf A}_b)_{ij} &=&  
\frac{
\sum_{l=1}^{N}
\sum_{\vec{q} = \vec{r},\vec{v}}\,
[{\sf Z}(t_{b+1}-t_b) \cdot{\sf L}_b\cdot\delta\vec{q}^{(i)}(\vec{r}_l,\vec{v}_l,t_b)]\cdot \delta\vec{q}^{(j)}(\vec{r}'_l,\vec{v}'_l,t_b) 
}{\sum_{l=1}^{N} \sum_{\vec{q} = \vec{r},\vec{v}}\, \delta\vec{q}^{(j)}\null(\vec{r}'_l,\vec{v}'_l,t_b)\cdot \delta\vec{q}^{(j)}(\vec{r}'_l,\vec{v}'_l,t_b)}\nonumber\\
&& \phantom{\sum_{l=1}^{N}\,\times}\times \exp[(- \mathrm{i} \vec{k}_i\cdot \vec{v}'_l+ \lambda_i ) \Delta t_b]~,
\label{eq:Aest}
\end{eqnarray}
Out of the total of $N$ particles, only two nearby particles are involved in each collision.
To leading order in $\bar{v}k_i/(\bar{\nu} N)$, their perturbations are the zero modes, which map onto themselves under collisions.
All the other particles are unaffected by the collision matrix.
As a result, after the short time-evolution after collision $b$ the modes are mapped exactly onto themselves, except for small terms for each particle due to the free flight, and larger collision terms for the two colliding particles.

The expressions in Eqs.~(\ref{eq:basis1}) and (\ref{eq:basis2}) may be inserted into Eq.~(\ref{eq:Aest}) to estimate the various terms.
Upon inserting the orthogonality of the basis and subtracting $\bar{\sf A}$, which approaches unity, one finds
\begin{eqnarray}
\epsilon {\sf X}^{ij}_b = O\left(\frac{\bar{v} k_i}{\bar{\nu} N}\right)~.
\end{eqnarray}
One may substitute this result into Eq.~(\ref{eq:wde}) while considering that
the eigenvalues $\kappa_i$ of the average tangent space evolution matrix $\bar{\sf A}$ are 
of order one.
In this way, it is found that the higher-order terms in the expansion are all of order $(\bar{v} k_i)^2/(\bar{\nu}^2 N)$.

In other words, the slowly growing and decaying modes couple sufficiently weakly to the other modes if $\bar{v} k/\bar{\nu} \ll 1$.
This is the case if the modes vary slowly in space, i.e., if the wavelength is smaller than the mean free path, as in Eq.~(\ref{eq:kklein}).
In this perturbation expansion all the higher-order corrections in Eq.~(\ref{eq:wde}) are of order $(\bar{v} k/\bar{\nu})^2$, one order of $\bar{v} k/\bar{\nu}$ smaller than the leading-order term.
The perturbation parameter needed for the weak-disorder expansion is therefore $\bar{v} k/\bar{\nu}$, and not $\bar{v} k/(\bar{\nu}N)$, because of the size of the disorder matrices.
As the wavenumber $k$ appears in the matrix ${\sf O}_b$, it appears in ${\sf A}_b$, and can appear in the perturbation parameter.

Note that the perturbation expansion can only be used to calculate the small Lyapunov exponents, when $\bar{v} k/\bar{\nu} < 1$.
If $k$ becomes too large, the remaining terms in Eq.~(\ref{eq:wde}) may not converge.

\section{\label{sec:comparison}Comparison and discussion}
Exactly the same approximations are used in both the Enskog approach (Sec.~\ref{sec:boltzmann}) and the random-matrix approach (Secs.~\ref{sec:rmt}, \ref{sec:equations}, and~\ref{sec:perturbationparameter}), namely the continuum limit, the Sto{\ss}zahlansatz, and the approximation of long wavelength, Eq.~(\ref{eq:kklein}).
The same results are found.
However, the method for arriving at these results is quite different, and the two approaches shed different light on the approximations and their validity.

In the derivation of the extended Enskog equation using kinetic theory, one starts with the approximations made in the standard Boltzmann and Enskog equations, namely the Sto{\ss}zahlansatz and the continuum limit, keeping in mind that the wavelength of the modes must be longer than the mean free path.
Next, the equations are multiplied by the first moments and integrated.

In the random-matrix approach, only a very general assumption of randomness is made at the beginning.
Then weak disorder is used, along with the approximation of long wavelength, followed by the Sto{\ss}zahlansatz, which is needed to simplify the equations for the basis.
And only then the continuum is limit taken.

As to the Sto{\ss}zahlansatz, in the Enskog approach it is used earlier, and the consequences of the Sto{\ss}zahlansatz are clearer.
In Ref.~\cite{onszelf} we showed that this approximation may not be sufficient for finding the Lyapunov exponents to leading order in the density.
From the kinetic-theory approach it is clear how to add correction terms to the equations for ring collisions and other similar trajectories.
In the random-matrix approach this is not so transparent, as the weak-disorder expansion theorem simply requires independence of the matrices.
It is not directly obvious how to extend the theorem to products of correlated matrices, or how to include ring-collisions in the equation for the basis sets for which the matrices become independent.
An exploration of the ring-collision terms within the framework of the kinetic-theory approach was made in Ref.~\cite{proefschrift}.

In the derivation of the generalised Enskog equation it is assumed, because the tangent space is linear, that the behaviour of the system can be described by the average behaviour of the first moments of the tangent space vectors.
This is equivalent to the requirement that the averages of the first moments of the matrix elements behave as if the matrices are independent.
Although an expansion in $\bar{v}k/\bar{\nu}$ of the type written down in Eq.~(\ref{eq:wde}) was not explicitly considered in the kinetic-theory approach, the same requirements of long wavelength and the Sto{\ss}zahlansatz were found to be needed.
Clearly, the original assumption made when deriving the generalised Enskog equation is not as trivial as it may seem.

The continuum limit is taken at the very end in the random-matrix approach.
Because of this, it is much more transparent than in the generalised Enskog equation how to add terms for finite-size effects to Eqs.~(\ref{eq:boltzr}) and (\ref{eq:boltzv}).
In particular, the longitudinal and transverse modes, and also modes with different values of $k$, are not perfectly orthogonal in systems with a finite number of particles.
Such nonorthogonality is directly related to fluctuations (in for instance the energy density) and its decay is related to the decay of fluctuations and correlation of position, energy, and momentum.
The average fluctuations of the inner products between modes with different wave numbers are zero, and therefore finite-size effects cannot contribute to the Lyapunov exponents to the leading order in $k$.
However, the relative size of these fluctuations is of order $1/\surd N$ and the total number of wave vectors is of order $N$.
There could be contributions to the higher-order terms on the right side of Eq.~(\ref{eq:wde}), which are of order $k^2/N$, leading to corrections to the Lyapunov exponents of order~$k^2$.

\section{\label{sec:conclusion}Conclusions}

The Lyapunov exponents close to zero can be related to Goldstone modes.
In this paper, I have shown that the equations for the small positive Lyapunov exponents found in Ref.~\cite{onszelf} by using Enskog theory can be derived from random-matrices by use of the weak-disorder expansion.
Precisely the same approximations are necessary in both approaches, namely the Sto{\ss}zahlansatz, the continuum limit, and the approximation of long wavelength.
In the random-matrix approach these approximations are needed in order to find a suitable description of the basis to meet the requirements of the weak-disorder-expansion theorem.
The approaches were thus found to be equivalent.

The different order in which the approximations are made, however, enables us to gain understanding into the consequences of the approximations and the uncertainties introduced by them.
In the kinetic-theory approach, it is more transparent how to add corrections for ring collisions and correlations, whereas in the random-matrix approach it is more clear how to add corrections for shorter wavelengths
and 
finite-size effects.
The latter may give insight into connections between the Lyapunov exponents and the decay of correlations and fluctuations.
Similar connections were suggested by Taniguchi and Morriss~\cite{tnm}.
Further investigation of the continuum limit and the validity of the weak-disorder expansion could perhaps also provide insight into the precise nature of the destruction of the Goldstone modes as the wavelength becomes too short compared to the mean free path.

In systems with soft potentials, the calculations described in this paper and in Ref.~\cite{onszelf} become more complicated.
Though the Goldstone modes are present in simulations of gases of soft particles, the degeneracy disappears~\cite{radons1,soft2,soft}.
This could be related to the fact that the zero modes are not as trivial in soft-potential systems as they are in system with hard-core interaction~\cite{proefschrift}.
Soft potentials can produce qualitative differences in the Lyapunov exponents, as has also been concluded in Refs.~\cite{lorentzpotentiaal,markusbeims}.

\end{document}